# An Efficient Fock Space Multi-reference Coupled Cluster Method based on Natural Orbitals: Theory, Implementation, and Benchmark

Soumi Haldar, and Achintya Kumar Dutta*

*Department of Chemistry, Indian Institute of Technology Bombay, Powai, Mumbai 400076.*

We present a natural orbital-based implementation of the intermediate Hamiltonian Fock space coupled-cluster method for (1,1) sector of Fock space. The use of natural orbital significantly reduces the computational cost and can automatically choose an appropriate set of active orbitals. The new method retains the charge transfer separability of the original intermediate Hamiltonian Fock space coupled-cluster method and gives excellent performance for valence, Rydberg, and charge-transfer excited states. It offers significant computational advantages over the popular equation of motion coupled cluster method for excited states dominated by single excitations.

*achintya@chem.iitb.ac.in

## I. Introduction:

The coupled cluster method has emerged as one of the most accurate and systematically improvable wave-function based method for ground and excited states in the last few decades[1]. The single reference coupled cluster is generally extended to excited states using the equation of motion (EOM) approach[2–7]. The equation of motion coupled cluster (EOM-CC) method is generally used in singles and doubles approximation (EOM-CCSD), which scales as $O(N^6)$ power of the basis set and has an $O(N^4)$ scaling storage requirement in the standard canonical implementation[2]. Although various computational strategies based on perturbative approximation[8–11], density fitting[12,13], semi-numerical approximation[14,15], use of local[16–18] and natural orbitals[19–26] have been described in the literature to reduce the computational scaling and/or storage requirement of the EOM-CCSD method. The ability to treat both dynamic and non-dynamic correlation within a conceptually simple framework and black-box nature of the EOM-CC method has made it extremely popular for chemical applications and is implemented in many quantum chemistry software packages. Extension of the EOM-CC method for triples (partial and full)[5,27–34] and higher-order corrections[35,36], analytic gradient[37], transition property calculation[2], first order[38] and second-order property[39] have also been achieved. The EOM-CC method gives identical results as that of the coupled-cluster linear response approach[40], although they are derived from very different theoretical viewpoints and is very similar to the symmetry adapted cluster CI (SAC-CI) approach of Nakatsuji and co-workers[41].

Alternatively, one can include the dynamic and non-dynamic correlation by using the multi-reference coupled cluster (MRCC) approach[27,42,43]. In MRCC methods, the dynamic correlation effect is included by the cluster amplitudes and non-dynamic correlation incorporated by diagonalization of the effective Hamiltonian. Depending upon the nature of the vacuum used, there can be two major flavors of MRCC methods. The Hilbert space or state universal MRCC approach[42,44] uses different vacuum for different cluster operators and is more suitable for the potential energy surfaces. The state-specific approaches to the Hilbert space coupled-cluster approach have also been defined[45]. On the other hand, the Fock space or valence universal MRCC[46–51] uses a common Fermi vacuum for all the cluster operators and more suitable for different energy calculations. In this approach, the standard single reference coupled cluster method serves as the zero-valence sector, and other reference determinants are generated by hierarchically adding quasi-particles until the desired sector of Fock space is reached. Difference energies like ionization potential[52,53], electron affinity[54,55], excitation energy[56–58], double ionization potential[59], double electron affinity[60], and triple electron affinity[61] can be calculated using the appropriate Fock-space MRCC (FSMRCC). There exists a third category of MRCC method, which uses the internally contracted coupled-cluster method (ic-MRCC)[62] based on Kutzelnigg-Mukherjee normal ordered ansatz[63].

In this work, we are primarily going to focus on Fock-space MRCC. The FSMRCC has also been successfully used for the calculation of energy difference[52–61] and potential energy surfaces[59,60,64]. Musial and co-workers have implemented the partial[61] and full[65,66] triples correction within the framework of FSMRCC. Other groups have also reported similar triples correction schemes for FSMRCC[53,67,68]. The triples correction schemes in FSMRCC for excitation energies have been shown to be computationally more efficient than the corresponding EOM-CC approaches, which makes them an extremely attractive option for excitation energy calculation[65]. Implementation of transition[69–72] and electrical properties[73] within the framework of FSMRCC has also been reported. Despite theoretical elegance and efficiency, the FSMRCC has never become the workhorse for routine excitation energy calculation. To the best of our knowledge, the only public domain software in which the FSMRCC is available is DIRAC[74,75] and EXP-T[76]. Many efficiency features for speeding up standard coupled cluster methods are not available for the FSMRCC implementation, e.g., treating the 4-virtual terms in the Atomic Orbital (AO) basis set, use of density fitting approximation. The accuracy of the FSMRCCSD for different kinds of excited states has not also been well benchmarked. In some sense, the implementation of the method has always remained in a pilot stage of development. The equally problematic thing is that the accuracy of the excitation energy in FSMRCC depends upon the used active orbitals. The accurate selection of active orbitals poses a significant barrier for the general users. This paper aims to describe a new natural orbital-based implementation of the FSMRCC method for excited states, which significantly reduces the computation cost and automatically selects an appropriate set of active orbitals for the calculation.

## II.  Theory and Computational Details:

The basic idea of FSMRCC theory is to obtain some selective eigenvalues of the Hamiltonian operator from the total eigenvalue spectrum. For this purpose, configuration space is divided into the model (P) and orthogonal spaces (Q) (See Figure 1). The projection operator for the model space is defined as[27]

$$\hat{P}_M^{(p,h)} = \sum_i \left| \Phi_i \right\rangle \left\langle \Phi_i \right| \tag{1}$$

$$\Phi_i = a_a^\dagger a_r \Phi \tag{2}$$

Where $\Phi$ is generally, but not necessarily a closed shell Hartree-Fock determinant. The r and a are labels of active occupied and unoccupied orbitals, respectively.

The number of active holes and particles determines the nature of the corresponding Fock space sector. For example, $P^{(1,0)}$ creates one active particle acting on an N electron wave-function and generates the N+1 electronic states i.e. singly electron attached states of the wave-function. Similarly, $P^{(0,1)}$ generates the model space for singly ionized states, $P^{(1,1)}$ generates the singly excited state, and so on. The projection operator for the orthogonal space is defined as

$$\hat{Q}_M^{(p,h)} = 1 - \hat{P}_M^{(p,h)} \tag{3}$$

if we consider a closed-shell system with n active occupied orbital levels then the number of "active" electrons is (2n-h+p). For (p,0) and (0,h) the model spaces are complete, for all other cases, we have the so-called quasi-complete model spaces[43]. The valence universal wave-operator introduces the dynamic correlation

$$\Omega = \{ e^{T+S} \} = e^T \{ e^{\tilde{S}^{(p,h)}} \} \tag{4}$$

The $W$ generates the exact wave-function from the model space. The cluster operator $\tilde{S}^{(p,h)}$ can destroy exactly p active particle and h active hole, in addition to creation of holes and particles. The braces in equation (4) indicate the normal ordering of cluster operators and $\tilde{S}^{(p,h)}$ is defined as[27]

$$\tilde{S}^{(p,h)} = \sum_{k=1}^p \sum_{l=1}^h \hat{S}^{(k,l)} \tag{5}$$

The $\hat{S}^{(k,l)}$ for particular sector subsumes all the lower sector Fock space amplitudes. The model space for the (0,0) sector consists of a single determinant (generally the Hartree-Fock determinant) $\mathsf{F}_0$ and the orthogonal space consists of all the excited determinants generated from $\mathsf{F}_0$ and is denoted as $\Phi^*$. Using this definition of P and Q space in the standard Fock-space block equation[27], one can obtain the following expression for the (0,0) sector

$$\left\langle \mathsf{F}^* \left| e^{-\hat{T}} H_N e^T \right| \mathsf{F} \right\rangle = 0 \tag{6}$$

$$E_{CC} = \left\langle \mathsf{F}_0 \left| e^{-\hat{T}} H_N e^{\hat{T}} \right| \mathsf{F}_0 \right\rangle \tag{7}$$

which is same as the standard ground state coupled cluster equations. The $\hat{H}_N = \hat{H} - E_{HF}$ and $E_{CC}$ is the ground state coupled cluster energy.

The similarity transformed Hamiltonian

$$\bar{H} = e^{-\hat{T}} \hat{H}_N e^{-\hat{T}} \tag{8}$$

has a block structure. In the singles and doubles truncation of $\hat{T}$, the similarity transformed Hamiltonian has the following form

$$\bar{H} = \begin{pmatrix} E_{CC} & \bar{H}_{0S} & \bar{H}_{0D} & 0 \\ 0 & \bar{H}_{SS} & \bar{H}_{ST} & \bar{H}_{ST} \\ 0 & \bar{H}_{DS} & \bar{H}_{DD} & \bar{H}_{DT} \\ \bar{H}_{T0} & \bar{H}_{TS} & \bar{H}_{TD} & \bar{H}_{TT} \end{pmatrix} \tag{9}$$

Where S, D, and T represent singly excited, doubly excited, and triply excited configurations, respectively.

Diagonalization of the similarity transformed Hamiltonian[77] will lead to an identical eigenvalue spectrum as that of $\hat{H}_N$ and it leads to the so-called equation of motion coupled cluster (EOM-CC) approach. By diagonalizing the $\bar{H}$ within the space of suitably chosen configurations, one can easily 'hitch-hike' through different sectors of Fock space[6] in an easy but not always cost-efficient way.

One can define a second similarity transformed Hamiltonian[78,79]

$$\hat{G} = e^{-\hat{X}} \bar{H} e^{\hat{X}} = e^{-\hat{X}} e^{-\hat{T}} H_N e^{\hat{T}} e^{\hat{X}} \tag{10}$$

The diagonalization of $\hat{G}$ will also lead to the same eigenvalues as that obtained from the diagonalization of $\bar{H}$ and $\hat{H}_N$ and as such, provides no additional advantage over diagonalization of $\bar{H}$. However, one can choose $\hat{X}$ such that it leads to a block diagonal structure[51,79] of $\hat{G}$ and one can obtain selective roots of $\bar{H}$ by diagonalizing only a subspace of $\hat{G}$.

There can be multiple ways of selecting the $\hat{X}$. We follow the recipe suggested by Meissner and co-workers[80].

$$\hat{X} = \left\{ e^{\hat{S}} - 1 \right\} P_M \tag{11}$$

Which on substitution to equation (10) gives

$$\hat{G} = \left( 1 - \hat{X} \right) \bar{H} \left( 1 + \hat{X} \right) \tag{12}$$

At this point, one can use the Malrieu's concept[81] of intermediate Hamiltonian. One can divide the entire configuration space into three subspaces interested in primary, intermediate and outer spaces with projection operators $P_M$, $P_I$ and $Q_0$ (See Figure 1) respectively. The intermediate space acts as a buffer between the primary and outer space.

If $\hat{X}$ satisfy the condition[80]

$$P_I \hat{G} P_M = 0 \tag{13}$$

Then one can extract the eigenvalues that correspond to the principal space just by diagonalization of the effective Hamiltonian

$$\hat{H}_{eff} = P_M \hat{G} P_M \tag{14}$$

One can split the similarity transformation in equation (10) as two sequential similarity transformations[80]

$$\hat{G} = e^{-\hat{X}} \bar{H} e^{\hat{X}} = e^{-\hat{Z}} e^{-\hat{Y}} \bar{H} e^{\hat{Y}} e^{\hat{X}} = e^{-\hat{Z}} \bar{G} e^{\hat{Z}} \tag{15}$$

Where

$$\hat{X} = \hat{Y} + \hat{Z} \tag{16}$$

$$\hat{Z} = P_I X P_M \tag{17}$$

$$Y = Q_0 X P_M \tag{18}$$

and

$$\bar{G} = e^{-\hat{Y}} \bar{H} e^{\hat{Y}} \tag{19}$$

The diagonalization of $\hat{G}$ and $\bar{G}$ will yield identical eigenvalues as they are related by a similarity transformation. One can obtain the m roots corresponding to primary space by diagonalization of the intermediate Hamiltonian[80]

$$\hat{H}_I = P_0 \bar{G} P_0 = P_0 \left(1 - \hat{Y}\right) \bar{H} \left(1 + \hat{Y}\right) P_0 \tag{20}$$

The rest of the eigenvalues $\hat{H}_I$ are arbitrary and are of little significance[80]. Equation (20) can be further simplified to

$$\hat{H}_I = P_0 \bar{H}(1 + \hat{Y}) P_0 = P_0 \bar{H} P_0 + P_0 \hat{Y} P_M \tag{21}$$

$\hat{Y}$ is zero for the (0,1) and (1,0) sector of the Fock space, and the corresponding intermediate Hamiltonian are

$$\hat{H}_I^{(0,1)} = P_0^{(0,1)} \bar{H} P_0^{(0,1)} \tag{22}$$

and

$$\hat{H}_I^{(1,0)} = P_0^{(1,0)} \bar{H} P_0^{(1,0)} \tag{23}$$

To diagonalize the intermediate Hamiltonian $H_I^{(0,1)}$ and $H_I^{(1,0)}$ is the same as solving the EOM-CC equation[46] for the principal peak of IP and EA problem. In this work, we are mostly interested in the (1,1) sector of Fock space, i.e., single excitation with respect to the Hartree-Fock determinant, and stick to singles and doubles truncation of the cluster amplitudes, which lead to

$$\hat{Y}^{(1,1)} = Q_0^{(1,1)} \left[ S_2^{(1,0)} + S_2^{(0,1)} + S_2^{(1,0)} S_1^{(0,1)} + S_2^{(0,1)} S_1^{(1,0)} + S_2^{(1,0)} S_2^{(0,1)} \right] P_M^{(1,1)} \tag{24}$$

The $S^{(1,0)}$ and $S^{(0,1)}$ can be extracted by putting intermediate Normalization of the converged IP and EA-EOM eigenvectors[46].

$$S = V V_0^{-1} \tag{25}$$

Where $V_0$ is the EOM-CC coefficient corresponding to the model space and $V$ contains the one corresponding to orthogonal space. The explicit expressions for the $\hat{S}$ amplitudes are as follows

$$S_m^{k'} = -\sum_\lambda r_k^\lambda W_m^\lambda \qquad (26)$$

$$S_{ji}^{bm} = -\sum_\lambda r_{ji}^{b\lambda} W_m^\lambda \qquad (27)$$

$$S_e^{c'} = \sum_\lambda r_\lambda^{c'} W_\lambda^e \qquad (28)$$

$$S_{je}^{ba} = \sum_\lambda r_{j\lambda}^{ba} W_\lambda^e \qquad (29)$$

Where $l$ is the number of EOM-CC root, m and e are active occupied and virtual respectively denoting the model space and $k'$, and $c'$ are the inactive occupied and inactive virtual orbitals. The transformation matrix

$$W = V_0^{-1} \qquad (30)$$

needs to be calculated numerically by inverting the $V_0$ matrix for the IP and EA case.

The intermediate Hamiltonian is then constructed using (19) and diagonalized in the space of singly excited configurations to obtain the excitation energies. For more details of Meissner's version of intermediate Hamiltonian, the readers are advised to go through the original formulation papers of intermediate Hamiltonian[47–50,80,82].

There are other variants of the intermediate Hamiltonian Fock space coupled cluster methods described in the literature. Kaldor and co-workers[83–86] have extended the intermediate Hamiltonian Fock space to relativistic domains. The eigenvalue independent partitioning scheme of Mukherjee and co-workers[87] gives an alternate formulation of the intermediate Hamiltonian Fock space coupled cluster,

If one considers a modified expression for $Y^{(1,1)}$

$$\hat{Y}'^{(1,1)} = Q_0^{(1,1)} \left[ S_2^{(1,0)} + S_2^{(0,1)} + S_2^{(1,0)} S_2^{(0,1)} \right] P_0^{(1,1)} \qquad (31)$$

and diagonalize the modified intermediate Hamiltonian

$$\hat{H}'_I = P_0 \bar{H} P_0 + P_0 \hat{Y} P_0 \qquad (32)$$

It becomes identical with the similarity transformed EOM-CCSD (STEOM-CCSD) method of Nooijen and co-workers[79]. The two methods lead to identical results when all the occupied and virtual orbitals are active.

Now, the IH-FSMR-CCSD approach, as described above, has two major drawbacks. First, the solution of the $\hat{T}$ amplitudes using equation (6) scales as iterative $O(N^6)$ of the basis set, which makes the calculation extremely expensive. Although Meissner and co-workers[88,89] explored some second-order approximation in the line of IH-FSMRCC, they were never rigorously benchmarked. Secondly, the accuracy of the excitation energy in the incomplete model space of the (1,1) sector depends on the size of the model space. There is no proper recipe for choosing a good model space apriority, which can be problematic for general users.

We present a new formulation using natural orbitals as a solution for both problems.

***Approximation of ground state with natural orbitals:***

The use of MP2 natural orbitals has emerged as an effective way of reducing the computational cost of the ground state coupled cluster calculations. Among the various flavors of natural orbital available, we have used the domain-based local pair natural orbital (DLPNO) framework of Neese and co-workers[90–98]. In this framework, the occupied space is expanded in terms of localized occupied orbitals, and the virtual space is expanded in terms of correlation domain using projected atomic orbitals (PAOs)[99]. The correlation domains are further truncated using pair natural orbitals (PNOs)[100], which are obtained by diagonalizing approximate pair densities calculated using MP2 method.

$$D^{i'j'}_{\tilde{\mu}_{ij}\tilde{\nu}_{ij}} = \sum_{\tilde{a}_{i'j'}} d_{\tilde{\mu}_{i'j'}\tilde{a}_{i'j'}} n_{\tilde{a}_{i'j'}} d_{\tilde{a}_{i'j'}\tilde{\nu}_{i'j'}} \qquad (33)$$

where $D^{i'j'}_{\tilde{\mu}_{ij}\tilde{\nu}_{ij}}$ is the pair density for pair $i'j'$ in the basis of localized occupied and virtual space made up of nonredundant PAOs $\tilde{\mu}_{i'j'}$. The $d_{\tilde{\mu}_{i'j'}\tilde{a}_{i'j'}}$ is the transformation matrix between the PAO and PNO basis and the $n_{\tilde{a}_{i'j'}}$ is the occupation number of the corresponding PNO. The coupled cluster equations are solved as a pair correlation problem and the total correlation energy is written as the sum of the pair energies ( $E_{i'j'}$ ) of all the pair.

$$E_{corr} = \sum_{i',j'} E_{i'j'}$$ (34)

Now, the pairs which have pair energy above a certain threshold (strong pair) are treated at the coupled-cluster level in much compact virtual space provided by the PNOs, and the rest of the amplitudes are approximated using the MP2 method. More details about the implementational of DLPNO-CCSD can be found in reference[101–105]. Related PNO based developments are also reported by the group of Werner[106–112], Hattig[25,113–119], and Kallay[23,120].

After the solution of the coupled cluster method, the coupled cluster amplitudes are back-transformed to the canonical basis. In the first step, the virtual space of the amplitudes are transformed from the PNO to canonical basis using three successive transformations from PNO ( $\tilde{a}_{i'j'}, \tilde{b}'_{i'j'}, ...$ ) to PAO ( $\tilde{\mu}, \tilde{\nu}, .....$ ), then to atomic orbital ( $m, n, .....$ ) and finally to molecular orbital ( $a, b, ...$ ) basis

$$d_{\tilde{\mu}\tilde{a}_{i'j'}} t^{i'j'}_{\tilde{a}_{i'j'}\tilde{b}_{i'j'}\tilde{\nu}} d_{\tilde{b}_{i'j'}\tilde{\nu}} = l_{\mu\tilde{\mu}} t^{i'j'}_{\tilde{\mu}\tilde{\nu}} l_{\tilde{\nu}\nu} = c_{a\mu} t^{i'j'}_{\mu\nu} c_{\nu b} = t^{i'j'}_{ab}$$ (35)

$$d_{\tilde{\mu}\tilde{a}_{i'}} t^{i'}_{\tilde{a}_{i'}} = l_{\mu\tilde{\mu}} t^{i'}_{\tilde{\mu}} = c_{a\mu} t^{i'}_{\mu} = t^{i'}_{a}$$ (36)

In the next step the occupied orbitals are transformed from local ( $i', j'$ ) to the canonical basis ( $i, j$ )

$$t^{ij}_{ab} = U_{ii'} t^{i'j'}_{ab} U_{j'j}$$

$$t^{i}_{a} = U_{ii'} t^{i'}_{a}$$

The transformation matrix U from the localized to canonical occupied orbitals are obtained by diagonalizing the occupied-occupied block of the Fock matrix in the localized basis ( $F'$ ).

$$F = U^{\dagger} F' U$$ (37)

Subsequently, an approximate similarity transformed Hamiltonian is constructed using the back-transformed coupled-cluster amplitudes.

$$\bar{H} \simeq \bar{H}^{bt} = e^{-\hat{T}^{bt}} H e^{\hat{T}^{bt}} \tag{38}$$

Neese and-co-workers[21] have introduced this approach of reducing the computation cost for the solution of ground state wave-function in the coupled cluster-based excited state method in the context of the EOM-CCSD and its similarity transformed version. However, the origin of the idea of treating a part of the amplitudes using coupled cluster method and the rest of the amplitudes using perturbation theory and subsequent perturbative truncation of the similarity transformed Hamiltonian can be traced back to earlier work by Nooijen[8,121], Sherrill[122], Stanton and Gauss[9]. The use of an extremely loose truncation threshold would lead to all the $t$ amplitudes being MP2 amplitudes and use of the corresponding similarity transformed Hamiltonian for the solution of FS-MRCC equation will lead to IH-FSMR-CCSD(2) method, similar to that of the EOM-CCSD(2) method of Stanton and Gauss[9]. On the other hand, the use of an extremely tight truncation threshold will lead to all the amplitudes being treated at the coupled-cluster level, and the resulting method will be the same as the standard IH-FSMR-CCSD. However, for all intermediate cases, one needs to have a closer look at the structure of the similarity transformed Hamiltonian constructed from the back-transformed amplitudes.

$$\bar{H}^{bt} = \begin{pmatrix} E_{CC}^{bt} & \bar{H}_{0S}^{bt} & \bar{H}_{0S}^{bt} & 0 \\ \hat{Z}_1 & \bar{H}_{SS}^{bt} & \bar{H}_{SD}^{bt} & \bar{H}_{ST}^{bt} \\ \hat{Z}_2 & \bar{H}_{DS}^{bt} & \bar{H}_{DD}^{bt} & \bar{H}_{DT}^{bt} \\ \bar{H}_{T0}^{bt} & \bar{H}_{TS}^{bt} & \bar{H}_{TD}^{bt} & \bar{H}_{TT}^{bt} \end{pmatrix} \tag{39}$$

The structure of the $\bar{H}^{bt}$ differs from the canonical $\bar{H}$ in equation (9) in the $\hat{Z}$ operators' presence, which gives the residual from the approximate solution to the CCSD equations, which is not zero because we are using approximate coupled cluster amplitudes. To solve this problem, we resort to a time-tested technique of neglecting the $\hat{Z}$. We define a projected similarity transformed Hamiltonian,

$$\bar{H}_0^{bt} = \begin{pmatrix} E_{CC}^{bt} & \bar{H}_{0S}^{bt} & \bar{H}_{0S}^{bt} & 0 \\ 0 & \bar{H}_{SS}^{bt} & \bar{H}_{SD}^{bt} & \bar{H}_{ST}^{bt} \\ 0 & \bar{H}_{DS}^{bt} & \bar{H}_{DD}^{bt} & \bar{H}_{DT}^{bt} \\ \bar{H}_{T0}^{bt} & \bar{H}_{TS}^{bt} & \bar{H}_{TD}^{bt} & \bar{H}_{TT}^{bt} \end{pmatrix} \tag{40}$$

for which the elements containing $\hat{Z}$ are explicitly removed. The $\bar{H}_0^{bt}$ can then be subsequently used to solve for the higher sectors of FSMRCC. The use of the projected similarity transformed Hamiltonian ensures the size-intensive excitation energy from the methods using back-transformed coupled-cluster amplitudes. It should be noted that $E_{CC}^{bt}$ is the coupled cluster energy obtained from the back-transformed coupled cluster amplitudes and is not the same as the standard DLPNO-CCSD energy.

### *Selection of the active orbitals with natural orbitals:*

The active orbital in the IH-FSMR-CCSD corresponds to the dynamic correlation of the quasi-particles, and therefore one needs to choose active orbitals for the (1,1) sector such that most of the dynamic correlation is already included. But it isn't easy to make such a choice a priory. However, the eigenvector of the $\bar{H}$ will contain the information about the quasi-particles, and one can use the corresponding natural orbitals to choose an appropriate set of active orbitals. The intermediate Hamiltonian can be subsequently built in the corresponding natural orbital basis. However, the solution of the full EE-EOM-CCSD equation before the solution of FSMR-CCSD for the selection of active orbitals will defeat the very purpose of using IH-FSMR-CCSD for excitation energy calculation. However, one can use the natural orbitals from an approximate method like CIS(D) or even CIS to determine the active orbitals.

After the generation of the back-transformed T amplitudes, a partial integral transformation has been performed, and CIS equations are solved. Subsequently, the state averaged CIS natural orbitals are calculated as follows

$$D_{ij}^{CIS} = \frac{1}{N} \sum_{I,a} R_a^i(I) R_a^j(I) \,\square \tag{41}$$

$$D_{ab}^{CIS} = \frac{1}{N} \sum_{I,i} R_a^i(I) R_b^i(I) \tag{42}$$

Where $R_a^i(I)$ is the CIS eigenvector corresponding to root $I$, and N is the number of CIS roots. In the present manuscript, we have chosen the number of CIS roots to be the same as that of the roots asked for the IH-FSMR-CCSD calculations.

The doubles correction part of the density was calculated as

$$D_{ij}^{D} = \frac{1}{N} \sum_{\lambda,kab} 2 R_{ab}^{ik}(\lambda) R_{ab}^{jk}(\lambda) - R_{ab}^{ik}(\lambda) R_{ba}^{jk}(\lambda) \tag{43}$$

$$D_{ab}^{D} = \frac{1}{N} \sum_{\lambda,ijc} 2 R_{ca}^{ij}(\lambda) R_{cb}^{ij}(\lambda) - R_{ca}^{ij}(\lambda) R_{bc}^{ij}(\lambda) \tag{44}$$

Where $R_{ab}^{ij}(\lambda)$ is the doubles part of CIS(D) vector for state $\lambda$ and is defined as

$$R_{ab}^{ij}(\lambda) = \frac{\sum_{c}\left[\langle ib|ca\rangle r_{c}^{j} + \langle jb|ca\rangle r_{c}^{i}\right] + \sum_{l}\left[\langle ji|ka\rangle r_{a}^{l} + \langle ji|kb\rangle r_{b}^{l}\right]}{\omega_{\lambda} + F_{ii} + F_{jj} - F_{aa} - F_{bb}} \tag{45}$$

Where $w_{\lambda}$ is the CIS excitation energy corresponding to state $\lambda$ and the total CIS(D) density is defined as

$$D_{ij}^{CIS(D)} = D_{ij}^{CIS} + D_{ij}^{D} \tag{46}$$

$$D_{ij}^{CIS(D)} = D_{ij}^{CIS} + D_{ij}^{D} \tag{47}$$

One can calculate the natural orbitals by separately diagonalized the occupied and virtual block of the state averaged CIS or CIS(D) density matrix.

$$V^{\dagger}DV = \hbar \tag{48}$$

The occupied and virtual orbitals up to a certain threshold of the natural orbital occupancy $\hbar$ have been chosen to be active. We have used a threshold of 0.001 after extensive testing. In the next step, the Fock matrix is transferred to the state-averaged natural orbital basis

$$\tilde{F} = VFV^{\dagger} \tag{49}$$

Subsequently, the active-active (AA) and inactive-inactive (II) block of the occupied and virtual Fock matrix is separately diagonalized. This gives a block diagonal structure of the Fock matrix.

$$F'' = W\tilde{F}W^{\dagger} = \begin{pmatrix} \varepsilon_{II}^{O} & F_{IA}^{''O} & 0 & 0 \\ F_{AI}^{''O} & \varepsilon_{AA}^{O} & 0 & 0 \\ 0 & 0 & \varepsilon_{AA}^{V} & F_{AI}^{''V} \\ 0 & 0 & F_{IA}^{''V} & \varepsilon_{II}^{V} \end{pmatrix} \tag{50}$$

Where $_e$ denotes the diagonal block of the Fock matrix $F_{IA}^{\prime\prime O}, F_{AI}^{\prime\prime O}$ and $F_{IA}^{\prime\prime V}, F_{AI}^{\prime\prime V}$ denotes of diagonal block of the occupied-occupied and virtual-virtual part of Fock matrices. The above transformation does not mix the occupied and virtual part of the Fock Matrix and Brillouin condition remains satisfied.

$$F_{IA} = \tilde{F}_{IA} = F''_{IA} = 0 \tag{51}$$

The back-transformed amplitudes are transformed to the state-averaged natural orbital basis

$$t_{a^*b^*}^{i^*j^*} = W_i^{i^*}W_{a^*}^{\tilde{a}}V_i^{\tilde{i}}V_{\tilde{a}}^{a}t_{ab}^{ij}V_b^{\tilde{b}}V_j^{\tilde{j}}W_{b^*}^{\tilde{b}}W_j^{j^*} \tag{52}$$

$$t_{a^*}^{i^*} = W_i^{i^*}W_{a^*}^{\tilde{a}}V_i^{\tilde{i}}V_{\tilde{a}}^{a}t_a^i \tag{53}$$

The Hartree-Fock coeffects are transformed to the natural orbital basis

$$C'' = CV^{\dagger}W^{\dagger}$$

The $C''$ is used to generate all the necessary integrals via RI transformation, and subsequent IH-FSMR-CCSD calculation on (0,1) and (1,0) and (1,1) sectors are performed in the natural orbital basis. The most expensive four external integrals are never generated and treated using the semi-numerical chain of sphere (COSX)☐approximation[15,123,124]. As the IH-FSMR-CCSD calculation on (0,1) and (1,0) sectors are the same as the IP and EA-EOM-CCSD method and the existing EOM-CCSD implementation has been reused. Ref [15] can be consulted for more details on the ORCA EOM-CCSD implementation and use of the chain of sphere approximation. Table 1 presents the formal scaling of different steps of the natural orbital-based IH-FSMR-CCSD method (IH-FSMR-CCSD-NO) and the approximation used to reduce computational cost. The construction of $\hat{H}_I$ and diagonalization of using Davidson iterative diagonalization scales only as O(N$^4$). However, the preceding steps of construction of $H_0^{bt}$ and solution of IP-EOM and EA-EOM CCSD calculations are of higher scaling. The most expensive step in the current implementation of IH-FSMR-CCSD-NO is the construction of $H_0^{bt}$ which scales as non-iterative O(N$^6$) power of the basis set. However, this step's formal scaling is independent of the number of roots calculated and can be of considerable advantage to the EOM-CCSD method when large number of roots are calculated. The explicit expressions for $\hat{H}_I$ intermediaries are presented in the appendix I. The use of state-average CIS or CIS(D) natural orbitals will not lead to any change in IP, and EA values for (1,0) and (0,1) sectors. However, the excitation energy value in the incomplete model space of (1,1) sector will differ between the canonical and state-average CIS or CIS(D) natural orbital basis. The ground state DLPNO truncation thresholds are going to affect the results for all the sectors. However, the use of back-transformed coupled cluster amplitudes has been shown to introduce negligible errors for EOM-CCSD calculations. Therefore, one needs to carefully benchmark the accuracy of the natural orbital-based IH-FSMR-CCSD method for the (1,1) sector.

It should be noted that the use of approximate natural orbitals for excited state coupled-cluster calculation is not new. Valeev and co-workers[125] have used state averaged CIS(D) natural orbitals to truncate the virtual space of EOM-CCSD calculation. Hattig and co-workers[24,25,117,119], Baudin *et al.*[22], and Kallay and co-workers[23], on the other hand, have used state-specific CIS(D) natural orbitals to calculate the truncated virtual space in coupled-cluster calculations. In this work, we will use the state-average natural orbitals to choose a suitable set of active orbitals for the IH-FSMR-CCSD calculations, not for truncating the virtual space itself. Although it can be very well be used for the second purpose, especially using state-average CIS(D) natural orbitals. However, such an attempt is outside the scope of the present study. The CIS natural orbitals were previously used to select the active orbitals in STEOM-CCSD[126] and CASSCF calculations[127]. In this work, we have tested the performance of both CIS and CIS(D) state-averaged density for the selection of active orbitals in IH-FSMR-CCSD method, and the resulting implementation will be denoted as IH-FSMR-CCSD-NO(CIS) and IH-FSMR-CCSD-NO(CIS(D)), respectively, for rest of the manuscript.

All the calculations are done with the IH-FSMR-CCSD-NO code implemented in the development version of ORCA[128].

### III. Result and discussion:

In this study, we focus on the excited state dominated by single excitation. The singly excited state may be broadly classified into three broad categories: valence, Rydberg, and charge-transfer states. We have tested the performance of the newly implemented IH-FSMR-CCSD-NO for all three kinds of excited states and compared the performance with the popular EOM-CCSD method.

### III.A Valence Excited States :

The performance of various coupled cluster-based methods for valence excited state is well benchmarked[15,21,126,129–131]. The Thiel test set[132] is generally used for benchmarking. We have calculated the performance of both IH-FSMR-CCSD-NO(CIS) and IH-FSMR-CCSD-NO(CIS(D)) for the Thiel test set using the TZVP basis set. The def2-TZVP/C auxiliary basis set has been used for the calculation. The CC3 method[133] has been taken as the benchmark, and the benchmark values, except for the DNA bases, were taken from the work of Nooijen and co-workers[129]. The CC3 benchmark values for the DNA bases were taken from the work of Szalay and co-workers[130]. The EOM-CCSD results are taken from the work of Nooijen and co-workers[129]. The statistical analysis of the error with respect to the benchmark CC3 values is provided in Table 2. The excited states with a % singles character greater than 87% are included in the analysis, and the individual excitation energies are provided in Table S1 and S2 of the supporting information.

The IH-FSMR-CCSD-NO(CIS) method shows a maximum absolute deviation (MAD) of 0.32 eV for the singlet state. The result is slightly better than the MAD of the EOM-CCSD method (0.39 eV) for the singlet state. The IH-FSMR-CCSD-NO(CIS(D)) method shows a MAD of 0.27 eV, which is lower than the IH-FSMR-CCSD-NO(CIS) and EOM-CCSD method. However, the mean absolute error (MAE) and root mean square deviation (RMSD) of the EOM-CCSD method are much higher than both the variants of the IH-FSMR-CCSD-NO. The result in the IH-FSMR-CCSD-NO(CIS) and the IH-FSMR-CCSD-NO(CIS(D)) are very similar although, the former shows a larger spread of the results (See Figure 2(a)).

The trends are slightly different for the triplet states. Both the IH-FSMR-CCSD-NO(CIS) and the IH-FSMR-CCSD-NO(CIS(D)) method gives slightly inferior performance to the EOM-CCSD method. The MAD in both the IH-FSMR-CCSD-NO is 0.69 eV, which is higher than the MAD of 0.52 eV in the EOM-CCSD method. The RMSD and MAE values of both IH-FSMR-CCSD-NO(CIS) and the IH-FSMR-CCSD-NO(CIS(D)) are slightly larger than that of the EOM-CCSD method. It is interesting to note that the EOM-CCSD method's performance is better for the triplet state, and the reverse is true for the IH-FSMR-CCSD-NO. Both the IH-FSMR-CCSD-NO(CIS) and the IH-FSMR-CCSD-NO(CIS(D)) method gives similar performance for the valence type triplet states (see Figure 2(b)). The reason for the slightly inferior IH-FSMR-CCSD-NO is not very clear to us at this point, but similar trends for triplet state have also been observed for the related STEOM-CCSD method[129].

### III.B Rydberg Excited States:

The Rydberg excited states involve excitation of the electron to the very diffuse virtual orbitals, and they closely mimic the ionized states of the molecules. The Rydberg excited states have been found[134] to be the 'Achilles heel' of many of the second-order approximate wave-function based methods like CC2, P-EOM-MBPT2,ADC(2)). Therefore, it will be interesting to benchmark the performance of IH-FS-MR-CCSD-NO for the Rydberg excited states. Although, Szalay and co-workers[135,136] have recently demonstrated that the use of spin-component scaling techniques can mitigate the problem of approximate wave-function-based methods for the Rydberg excited states.

We have used the Waterloo test set of Nooijen and co-workers[137] for benchmarking the performance of IH-FS-MR-CCSD-NO methods. The Ahlrich's-TZVP basis set[138] was used supplemented with an additional 5s, 4p, and 4d functions added to a ghost atom placed at the center of symmetry were used for the calculations. The auxiliary basis functions used for the calculations are generated using the AUTOAUX[139] utility of ORCA. The COSX approximation[15,123,124] is used to calculate the term involving 4-external integrals in the EA-EOM-CCSD step of the IH-FS-MR-CCSD-NO method. The default grid setting for the radial grid (2.68 ) is known to give convergence issues in the EA-EOM-CCSD iterations when basis sets with defuse function were used. Therefore, a radial grid of 7 has been used for the Ryderg excited states in the IH-FS-MR-CCSD-NO method. The angular grid has been kept at the default value of 1. The CC3 results have been used as the benchmark for the singlet state, and the EOM-CCSD(T) method of Bartlett and co-workers[33] has been used as the benchmark for the triplet states. All the CC3, EOM-CCSD, and EOM-CCSD(T) results were taken from reference[129,140] Table 3 presents the statistical analysis of the error of the singlet and triplet states in IH-FS-MR-CCSD-NO(CIS) and IH-FS-MR-CCSD-NO(CIS(D)) along with the EOM-CCSD results for the corresponding states. The singlet and the triplet states' individual excitation energies are presented in Table S3 and S4 of the supporting information.

It can be seen that both the IH-FS-MR-CCSD-NO(CIS) and IH-FS-MR-CCSD-NO(CIS(D)) methods show quite similar performance with respect to the CC3 benchmark results, with the latter showing a slightly more extensive spread of errors (See Figure 3(a)). The MAD for the two methods is 0.28 eV and 0.33 eV, respectively, which is slightly better than the EOM-CCSD method. The RMSD error of IH-FS-MR-CCSD-NO(CIS) and IH-FS-MR-CCSD-NO(CIS(D)) is 0.09 eV and 0.10 eV, respectively, which is in the same order as the corresponding EOM-CCSD method. It is interesting to note that the performance of the EOM-CCSD method for singlet Rydberg state significantly improves as compared to their performance for valence excited states with MAE, and RMSD error becomes almost half of what is observed for the valence excited state. The performance of both IH-FS-MR-CCSD-NO(CIS) and IH-FS-MR-CCSD-NO(CIS(D)) method, on the other hand, is quite similar to their corresponding performance for valence excited states.

The IH-FS-MR-CCSD-NO(CIS) and IH-FS-MR-CCSD-NO(CIS(D)) methods show quite similar performance to one another for Rydberg triplet states (See Figure 3(b)). Both the method shows an MAE of 0.05 eV and RMSD error of 0.07 eV with respect to the benchmark EOM-CCSD(T) results. A direct comparison of the IH-FS-MR-CCSD-NO methods for the Rydberg and the valence triplet states is not possible as the errors are calibrated against different benchmark methods. However, the performance of The IH-FS-MR-CCSD-NO(CIS) and IH-FS-MR-CCSD-NO(CIS(D)) method improves from the valence to Rydberg state. The results for the Rydberg triplet states are comparable to the corresponding EOM-CCSD values.

### III.C Charge-Transfer Excited States :

The charge transfer separability, even at the CCSD approximation, is one of the strong points of the FSMRCC methods. The inter-fragment charge transfer excitation energy between two fragments separated at infinite distance in IH-FSMR-CCSD methods is the same as the sum of the donor fragment's ionization energy, the electron affinity of the acceptor fragment, and the long-range Coulomb interaction between two charged fragments and follows the relation

$$EE = IP + EA - \frac{e^2}{R}$$

The standard EOM-CC method shows a finite charge-transfer separability error in the singles and doubles approximation. One needs to include triples to get the correct charge transfer separability in the excitation energy of the standard EOMCC method. The readers can look at ref[141] by Musial and Bartlett to have a more elaborate discussion on the charge transfer separability in EOM-CC and FSMRCC methods. Now, it is desirable that natural orbital-based implementation of IH-FSMR-CCSD retains the original charge transfer separability of the standard canonical IH-FSMR-CCSD method. To investigate the charge transfer separability behavior of the IH-FS-MR-CCSD-NO(CIS) and IH-FS-MR-CCSD-NO(CIS(D)) method, we have calculated the excitation energy of the Be-$C_2$ model system. Musial and Bartlett used the same system to test the charge transfer separability of canonical IH-FSMR-CCSD method[65,141]. The model system was also previously used by Nooijen and Bartlet[79] to explore the charge transfer separability of the STEOM-CCSD method. The model system is constructed by placing the $C_2$ unit perpendicular to the axis connecting the Be and midpoint of the $C_2$ unit, as displayed in Figure 4. The cc-pVDZ basis set and the cc-pVDZ/C auxiliary basis set is used for the calculation, and the results for different internuclear separations are presented in Table 4. It can be seen that both IH-FS-MR-CCSD-NO(CIS) and IH-FS-MR-CCSD-NO(CIS(D)) method shows correct asymptotic $R^{-1}$ behavior barring tiny numerical noises. The COSX approximation used for the calculation of the term containing 4-external integral in EA-EOM-CCSD causes this numerical noise. One can completely remove the numerical noise by increasing the radial grid to a large value. However, that deemed to be unnecessary for all practical purposes. The EOM-CCSD method shows a charge separability error of 0.06453 eV at the internuclear separation of 10000Å between Be and $C_2$. The excitation energy in EOM-CCSD seems to overestimate from the correct asymptotic limit value.

To understand the practical implication of the charge transfer separability behavior, we have used the newly developed charge-transfer test set of Szalay and co-workers[142], consisting of dimers with low-energy CT states. The cc-pVDZ basis set and the cc-pVDZ/C auxiliary basis set have been used for the calculation and the EOM-CCSDT-3 results is used as the benchmark. All the EOM-CCSDT-3 and EOM-CCSD energies are taken from reference. It can be seen that IH-FS-MR-CCSD-NO(CIS) and IH-FS-MR-CCSD-NO(CIS(D)) give excellent performance for the charge transfer states with a MAD of 0.17 eV and 0.14 eV, respectively. The EOM-CCSD method systematically overestimates the results similar to the trend observed for the Be-$C_2$ model system with a mean error (ME) of 0.30 eV. On the other hand, the performance of the IH-FS-MR-CCSD-NO(CIS) and IH-FS-MR-CCSD-NO(CIS(D)) method is quite similar to that of that observed for valence and Rydberg state, with the former showing a little higher spread (See Figure 5).

Unfortunately, there are no benchmark values available for the CT test set for the triplet states.

**III.D Efficiency Consideration:**

The IH-FS-MR-CCSD-NO(CIS) and IH-FS-MR-CCSD-NO(CIS(D)) have almost similar accuracy. Therefore we have considered the IH-FS-MR-CCSD-NO(CIS) to benchmark the computational efficiency.

To investigate the efficiency of IH-FS-MR-CCSD-NO(CIS) in comparison to the canonical EOM-CCSD and IH-FS-MR-CCSD method, we have calculated the excitation energies corresponding to a series of water clusters (($H_2O$)$_n$,n=1,8). Total eight roots have been calculated for each cluster, and a def2-TZVP basis set has been used. The number of active holes and active particles in canonical IH-FSMR-CCSD are taken to be the same as those chosen by the automatic active orbital selection scheme in IH-FS-MR-CCSD-NO(CIS) method. Figure 6 shows that the computation time of the EOM-CCSD increases much more steeply with the number of water molecules than the IH-FSMR-CCSD, even in the canonical implementation. The IH-FS-MR-CCSD-NO(CIS) is much more efficient than the canonical IH-FS-MR-CCSD and the corresponding computational time rises much less steeply than the canonical IH-FS-MR-CCSD and EOM-CCSD.

To demonstrate the applicability of the IH-FS-MR-CCSD-NO(CIS) method beyond small molecules, we have calculated the excitation energies of ZMSO2M-14TPA (Z-methylsulfonylpropenyltriazolyl-triphenylamine) chromophore, which contains 53 atoms and 226 electrons. The same molecule was used as a test case by Plasser in his Theodore[143] implementation paper. The def2-TZVP basis set has been used, which leads to 1099 basis functions. Total 12 roots were calculated, and C1 symmetry has been used. This is one of the biggest FSMR-CCSD calculations ever performed. The automatic active orbital selection scheme has chosen 8 holes and 9 particles as active. The total calculation took 153 hours 45 minutes, of which only 6 minutes is spent in the Hartree-Fock calculation. Total 5 hours 41 minutes has been spent in the ground state DLPNO-CCSD calculation, and generation of integrals in the state averaged CIS-NO basis. The IP and EA-EOM-CC calculation took 94 hours 59 minutes. Total 52 hours 51 minutes has been spent in the construction of intermediate Hamiltonian and 14 minutes in the diagonalization.

Figure 7 presents the natural transition orbitals corresponding to the brightest state of ZMSO2M-14TPA. The transition densities are calculated using a CIS-like approximation with the $\hat{H}_I$ eigenvectors. The excitation energy corresponding to the brightest state is 3.86 eV. The ADC(2) results reported by Plasser[143] for the same state in the def2-SV(P) basis set is 3.95 eV, which is in the same range of our IH-FS-MR-CCSD-NO result.

### IV. Conclusions :

We present the formulation and implementation of a natural orbital-based intermediate Hamiltonian Fock coupled-cluster method for (1,1) sector of Fock space. The method uses the pair natural orbitals (PNO) for the (0,0) sector and state-averaged CIS/CIS(D) natural orbitals for the higher sectors. Natural orbitals decrease the computational cost of the calculations and can automatically choose appropriate active orbitals. We have benchmarked the performance of using both CIS and CIS(D) natural orbitals for valence, Rydberg, and charge transfer excited-states. CIS and CIS(D) natural orbitals' performance has been found to be quite similar for all kinds of excited states, which shows the results in IH-FSMR-CCSD are relatively insensitive to the precise definition of active orbitals. The IH-FSMR-CCSD-NO(CIS) and IH-FSMR-CCSD-NO(CIS(D)) method give comparable performance to the EOM-CCSD method for Rydberg singlet at a much lower computational cost. Both the IH-FSMR-CCSD-NO methods give better performance than the EOM-CCSD method for the valence and charge-transfer singlet states. The advantage is especially noticeable for the charge-transfer states where lack of charge-transfer separability of the EOM-CCSD method results in larger errors. The natural orbital-based implementation retain charge-transfer separability of the canonical IH-FSMR-CCSD method and gives uniform accuracy for valence, Rydberg, and CT states. The IH-FSMR-CCSD-NO methods give a slightly inferior performance for valence triplet states. However, the performance is comparable to the EOM-CCSD method for Rydberg triplet states.

Computationally The IH-FSMR-CCSD is much more efficient than the canonical IH-FSMR-CCSD or EOM-CCSD. The present implementation's efficiency is demonstrated by calculating the excitation energy corresponding to the brightest state of ZMSO2M-14TPA chromphore. The routine chemical application of these newly implemented IH-FS-MR-CCSD-NO will require the implementation of transition moment and other associated properties. The work is in progress towards that direction.

### Supporting Information:

All the excitation energies, the geometry of water clusters, the geometry and excitation energies of ZMSO2M-14TPA are provided in the supporting information.

**Acknowledgment**

The authors acknowledge the support from the IIT Bombay, IIT Bombay Seed Grant project, DST-Inspire Faculty Fellowship for financial support, IIT Bombay super computational facility, and C-DAC Supercomputing resources (PARAM Yuva-II, Param Bramha) for computational time.

Conflict of interest

The authors declare no competing financial interest.

**Appendix I:**

The sigma vector for

<u>Triplet state</u>

$$\sigma_i^a = \tilde{\bar{F}}_a^b r_i^a + \bar{\bar{F}}_a^b r_m^a \delta_{mi} - \tilde{\bar{F}}_i^{\,j} r_i^a - \bar{\bar{F}}_i^{\,j} r_j^e \delta_{ea} - \bar{\bar{J}}_{ab}^{\,ij} r_j^b$$

 <u>Singlet state</u>

$$\sigma_i^a = \tilde{\bar{F}}_a^b r_i^a + \bar{\bar{F}}_a^b r_m^a \delta_{mi} - \tilde{\bar{F}}_i^{\,j} r_i^a - \bar{\bar{F}}_i^{\,j} r_j^e \delta_{ea} + 2 \bar{\bar{K}}_{ab}^{\,ij} r_j^b - \bar{\bar{J}}_{ab}^{\,ij} r_j^b$$

where

$$\bar{\bar{F}}_m^i = \tilde{F}_k^b \tilde{S}_{ik}^{mb} - \hat{g}_{ib}^{kl} \tilde{S}_{kl}^{mb}$$

$$\bar{\bar{F}}_e^a = \tilde{F}_k^c \tilde{S}_{ek}^{ac} - \hat{g}_{cd}^{ld} \tilde{S}_{el}^{cd}$$

$$\bar{\bar{J}}_{ab}^{ij} = \tilde{J}_{ab}^{ij} + u_{ae}^{im} \delta_{mj} \delta_{eb} + u_{ae}^{im} \delta_{mj} \delta_{eb} + u_{ae}^{im} \delta_{mj} \delta_{eb}$$

$$\tilde{\bar{K}}_{ab}^{ij} = \tilde{K}_{ab}^{ij} + \tilde{u}_{ae}^{im} \delta_{mj} \delta_{eb} + \tilde{u}_{ae}^{im} \delta_{mj} \delta_{eb} + \tilde{u}_{ae}^{im} \delta_{mj} \delta_{eb}$$

$$\tilde{S}_{ij}^{mb} = 2S_{ij}^{mb} - S_{ji}^{mb}$$

$$\tilde{S}_{ej}^{ab} = 2S_{ej}^{ab} - S_{ej}^{ba}$$

$$u_m^i = \tilde{F}_k^b \tilde{S}_{ik}^{mb} - \hat{g}_{ib}^{kl} \tilde{S}_{kl}^{mb}$$

$$u_{ae} = \tilde{F}_k^c \tilde{S}_{ek}^{ac} - \hat{g}_{cd}^{ld} \tilde{S}_{el}^{cd}$$

$$u_m^c = -g_{cd}^{kl} S_{kl}^{md}$$

$$u_k^e = g_{cd}^{kl} \tilde{S}_{el}^{cd}$$

$$u_{ae}^{im} = \tilde{F}_k^e S_{ik}^{ma} + \hat{g}_{de}^{la} \tilde{S}_{il}^{md} - \hat{g}_{ed}^{la} S_{il}^{md} + \hat{g}_{ie}^{kl} S_{kl}^{ma}$$

$$u_{ae}^{im} = \tilde{F}_m^c S_{ei}^{ac} + \hat{g}_{id}^{ml} \tilde{S}_{el}^{ad} - \hat{g}_{id}^{lm} S_{el}^{ad} + \hat{g}_{cd}^{ma} S_{ei}^{cd}$$

$$u_{ae}^{im} = u_m^c S_{ei}^{ac} - u_e^k S_{ma}^{ik} + u_{il}^{mc} \tilde{S}_{el}^{ad} - u_{kd}^{im} S_{ek}^{ad} + \tilde{u}_{kd}^{im} S_{ek}^{da} + u_{ie}^{kl} S_{lk}^{ma} - u_{ae}^{ik'} S_m^{k'} + u_{ac'}^{im} S_e^{c'} + u_i^m S_e^{a'} \delta_{a'a} + u_e^a s_m^{k'} \delta_{k'i}$$

$$u_{il}^{mc} = g_{cd}^{kl} \tilde{S}_{mc}^{ik}$$

$$u_{kd}^{im} = g_{dc}^{kl} \tilde{S}_{mc}^{il}$$

$$\tilde{u}_{kd}^{im} = g_{dc}^{kl} S_{mc}^{il}$$

$$u_{ie}^{kl} = g_{cd}^{kl} S_{ei}^{cd}$$

$$\tilde{u}_{ab}^{im} = \hat{g}_{ib}^{kl} S_{lk}^{ma} - \hat{g}_{ka}^{bc} S_{ki}^{mc} + \tilde{F}_k^b S_{ki}^{ma}$$

$$\tilde{u}_{ae}^{ij} = \bar{F}_k^e S_{ie}^{ac} + \hat{g}_{ja}^{cd} S_{ei}^{cd} - \hat{g}_{ic}^{kj} S_{ke}^{ac}$$

$$\tilde{u}_{ae}^{im} = u_m^c S_{ei}^{ca} - u_e^k S_{ma}^{ki} + \tilde{u}_{kd}^{im} S_{ke}^{ad} + \tilde{u}_{ie}^{kl} S_{lk}^{ma} + \tilde{u}_{ab'}^{im} S_e^{b'} - \tilde{u}_{ae}^{ij'} S_m^{j'}$$

$$\tilde{u}_{kd}^{im} = g_{cd}^{kl} S_{jc}^{li}$$

$$\tilde{u}_{ie}^{kl} = g_{cd}^{kl} S_{ib}^{cd}$$

In the above equations,  I,j,k,l and a,b,c,d denotes occupied and virtual orbitals, respectively. The m and e denote active indices of the hole and particle type, respectively, while a prime denotes a restriction to orbitals that are not active. The Einstein summation convention has been used for all the expressions.

The $g_{cd}^{kl}$ are standard two-electron integrals. The $\tilde{J}$, $\tilde{K}$ and $\hat{g}$ are $\bar{H}_0^{bt}$ intermediates, the explicit programable expressions for them can be found in ref[15]                                                                 □

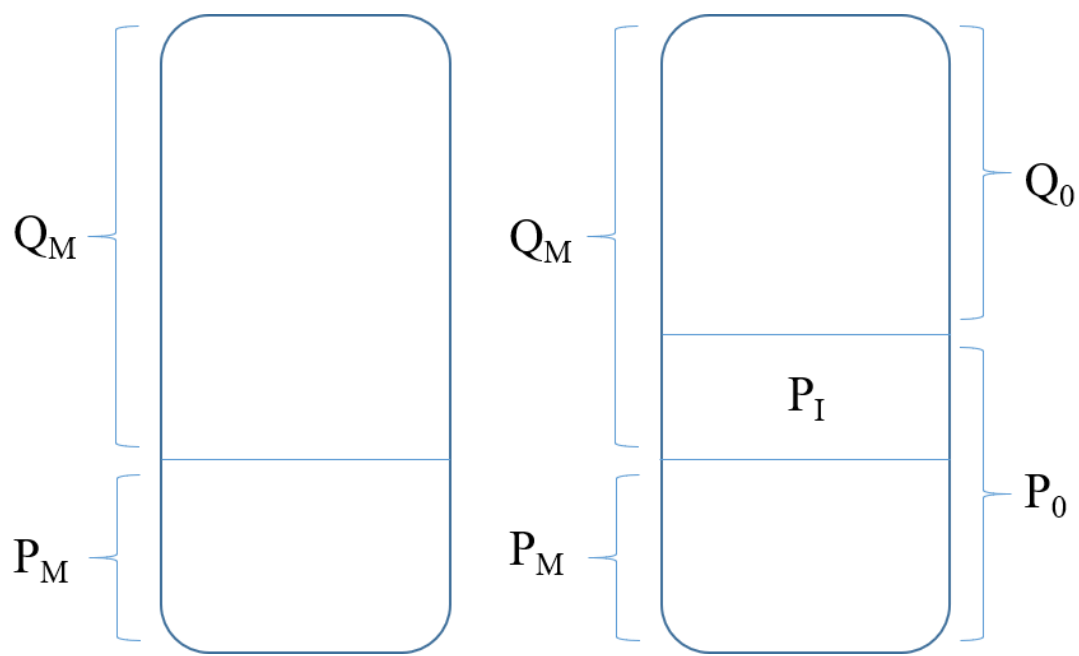

*Figure 1: The schematic depiction of the Model Space for effective and Intermediate Hamiltonian Formulation of FSMRCC*

*Table 1: The formal scaling and the approximation used in the different steps of the IH-FSMR-CCSD-NO method*

| Step | Formal Scaling | Apprxomtion used |
|---|---|---|
| Ground state CCSD calculation | Iter near-linear | DLPNO |
| Back transformation of CC amplitudes and state natural orbital construction | Non-iter $O(N^5)$ | RI. |
| Construction of $\bar{H}_0^{bt}$ | Non-iter $O(N^6)$/ iter $O(N^5)$ | State-average CIS/CIS(D) NO |
| Solution of IP and EA equations | Iter $O(N^5)$ | COSX for EA, State-average CIS/CIS(D) NO |
| Construction of $\hat{H}_l$ | Non-iter $O(N^4)$ | State-average CIS/CIS(D) NO |
| Diagonalization of $\hat{H}_l$ | Non-iter $O(N^4)$ | State-average CIS/CIS(D) NO |

*Table 2: The statistical analysis[a] of results for valence excited states (in eV).*

| | IH-FSMR-CCSD-NO(CIS) Singlet/triplet | IH-FSMR-CCSD-NO(CIS(D)) Singlet/triplet | EOM-CCSD Singlet/triplet |
|---|---|---|---|
| MAD | 0.32/0.69 | 0.27/0.69 | 0.39/0.51 |
| ME | -0.03/-0.13 | 0.00/-0.15 | 0.20/0.05 |
| MAE | 0.09/0.17 | 0.08/0.18 | 0.20/0.11 |
| RMSD | 0.12/0.22 | 0.10/0.23 | 0.22/0.16 |
| STD | 0.11/0.17 | 0.10/0.17 | 0.10/0.15 |

*[a]The statistical quantities are abbreviated as maximum absolute deviation (MAD), mean error (ME), mean absolute error (MAE), root mean square deviation (RMSD), and standard deviation (STD).*

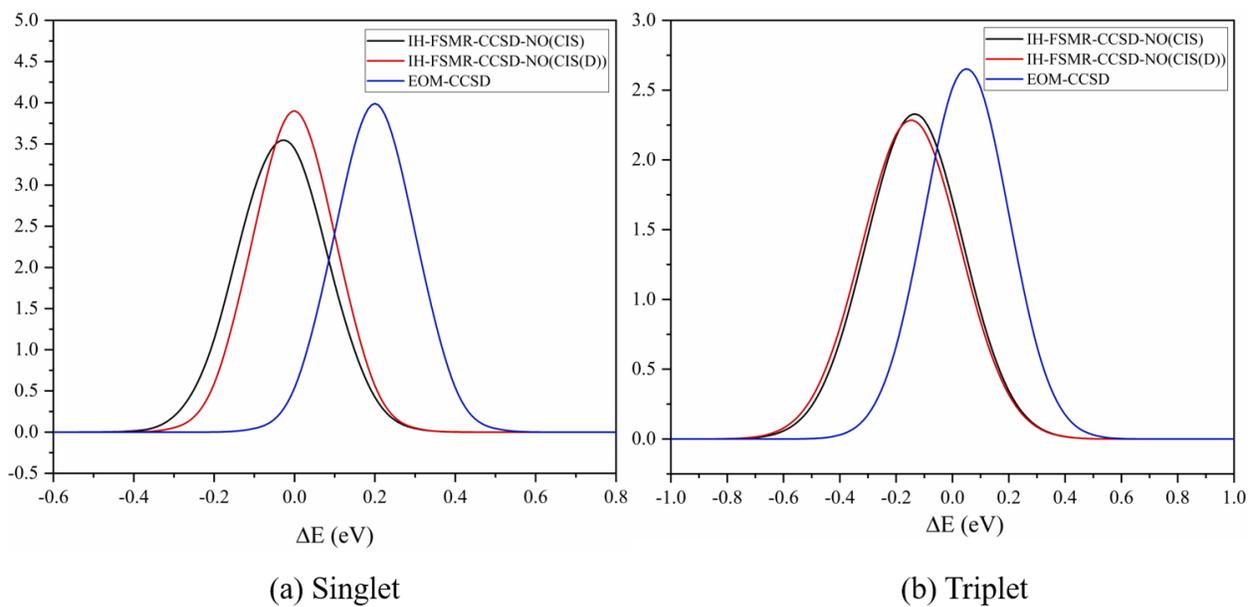

(a) Singlet                              (b) Triplet

*Figure 2: Error distribution plot corresponding to the (a)singlet and (b) triplet valence excited states.*

*Table 3:  The statistical analysis[a] of results  for Rydberg excited states (in eV)*

|  | IH-FSMR-CCSD-NO(CIS) Singlet/triplet | IH-FSMR-CCSD-NO(CIS(D)) Singlet/triplet | EOM-CCSD Singlet/triplet |
|---|---|---|---|
| MAD | 0.28/0.30 | 0.328/0.36 | 0.39/0.42 |
| ME | 0.06/0.03 | 0.067/0.03 | 0.08/0.05 |
| MAE | 0.07/0.05 | 0.08/0.05 | 0.08/0.05 |
| RMSD | 0.09/0.07 | 0.10/0.07 | 0.10/0.07 |
| STD | 0.06/0.06 | 0.07/0.06 | 0.06/0.05 |

[a]The statistical quantities are abbreviated as maximum absolute deviation (MAD), mean error (ME), mean absolute error (MAE), root mean square deviation (RMSD), and standard deviation (STD).

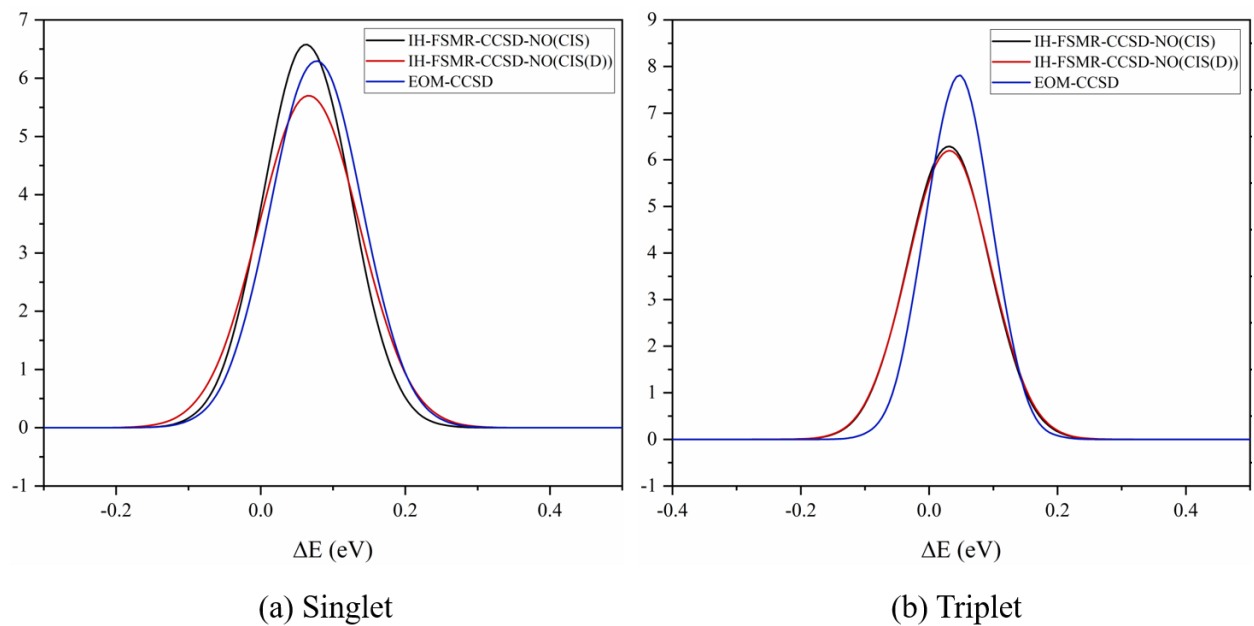

*Figure 3: Error distribution plot corresponding to the (a) singlet and (b)triplet Rydberg excited states.*

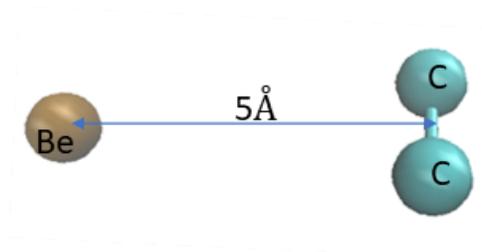

*Figure 4: Be–C$_2$ model complex, separated at 5 Å.*

*Table 4: Behavior of a charge transfer excitation in Be–$C_2$ model system with increasing distance between fragments (R)*

| | $-e^2R^{-1}$ (eV) | IH-FSMR-CCSD-NO(CIS) (default radial grid for COSX) | IH-FSMR-CCSD-NO(CIS) (radial grid 20 for COSX) | IH-FSMR-CCSD-NO((CIS) (D)) (default radial grid for COSX) | EOM-CCSD |
|---|---|---|---|---|---|
| R (Å) | | | | | |
| 5 | -2.8800 | -2.8111 | -2.8108 | -2.9195 | -2.8797 |
| 8 | -1.8000 | -1.7835 | -1.7833 | -1.7833 | -1.7187 |
| 10 | -1.4400 | -1.4320 | -1.4318 | -1.4318 | -1.3675 |
| 100 | -0.1440 | -0.1440 | -0.1440 | -0.1440 | -0.0793 |
| 1000 | -0.0144 | -0.0144 | -0.0144 | -0.0144 | 0.0501 |
| 10000 | -0.0014 | -0.0015 | -0.0014 | -0.0014 | 0.0631 |

[EE-(IP+EA)] (eV)

*Table 5:The statistical analysis of results for singlet charge transfer states (in eV)*

|        | IH-FSMR-CCSD-NO(CIS) | IH-FSMR-CCSD-NO(CIS(D)) | EOM-CCSD |
|--------|----------------------|-------------------------|----------|
| MAD    | 0.17                 | 0.14                    | 0.44     |
| ME     | -0.04                | -0.03                   | 0.30     |
| MAE    | 0.07                 | 0.06                    | 0.30     |
| RMSD   | 0.09                 | 0.07                    | 0.31     |
| STD    | 0.08                 | 0.07                    | 0.08     |

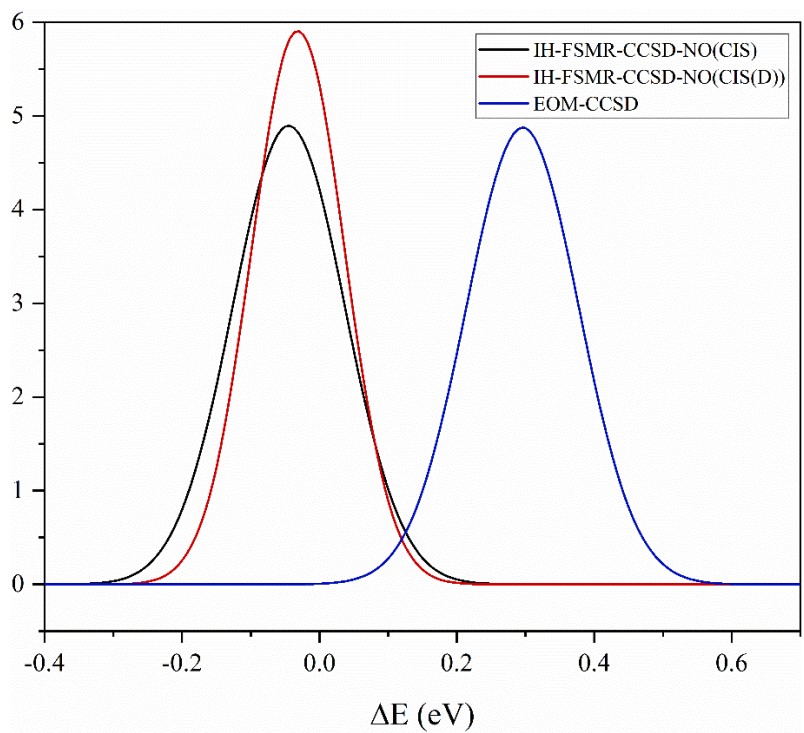

*Figure 5: Error distribution plot corresponding to the charge transfer excited states.*

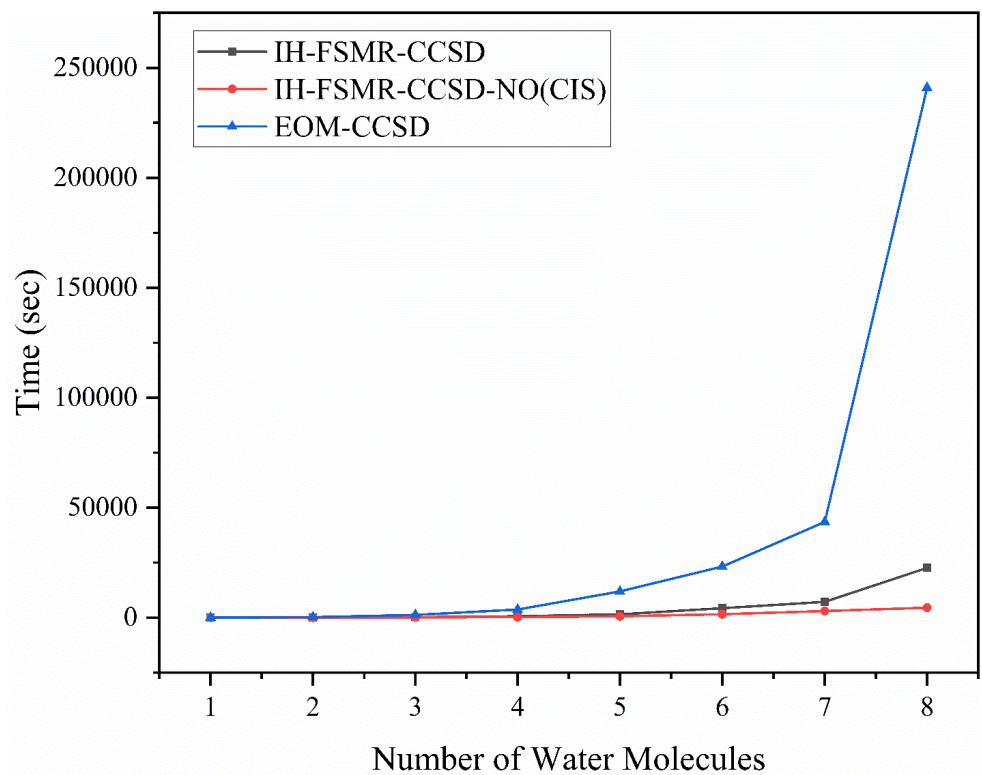

*Figure 6: The times taken by different methods in the correlation calculation for a series of water clusters ($(H_2O)_n$, n=1,8).*

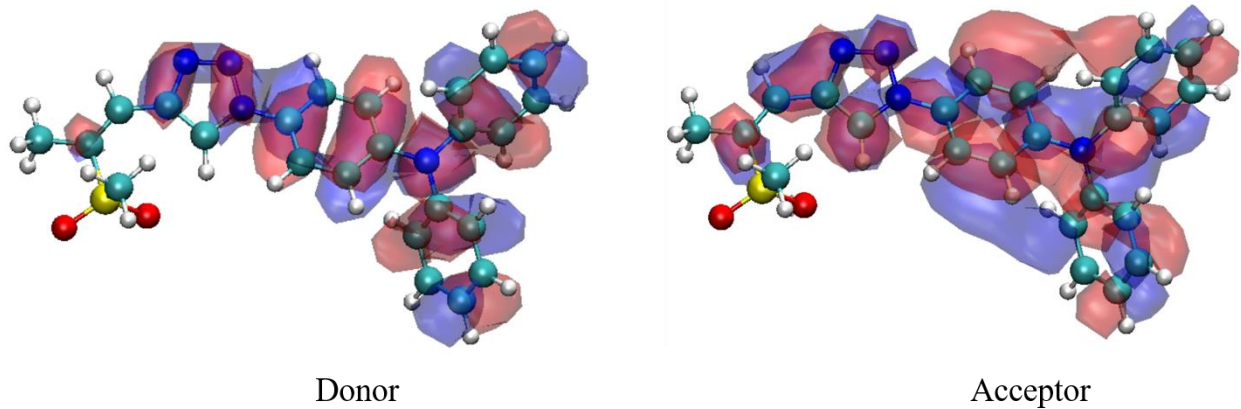

Donor                               Acceptor

*Figure 7: Natural transition orbitals corresponding to the brightest state of ZMSO2M-14TPA*
*chromphore*